\documentclass[letter]{aa}

\usepackage{amsmath}
\usepackage{graphicx}
\usepackage{graphics}
\usepackage{epsfig}
\usepackage{txfonts}
\usepackage[colorlinks=true,linkcolor=blue,citecolor=blue,urlcolor=blue]{hyperref}
\begin{document}

\title{{\it Gaia}-assisted selection of a quasar reddened by
dust in an extremely-strong damped Lyman-$\alpha$ absorber
at $z=2.226$}

\titlerunning{A QSO reddened by a dusty DLA}

\author{
S.~J.~Geier\inst{1,2},
K.~E.~Heintz\inst{3},
J.~P.~U.~Fynbo\inst{4,5},
C.~Ledoux\inst{6},
L.~Christensen\inst{7},
P.~Jakobsson\inst{3},
J.-K.~Krogager\inst{8},
B.~Milvang-Jensen\inst{4,5},
P.~M\o ller\inst{9},
P.~Noterdaeme\inst{8}
}
\institute{
Instituto de Astrof{\'i}sica de Canarias, V{\'i}a L{\'a}ctea, s/n, 38205, La Laguna, Tenerife, Spain --
\email{stefan.geier@gtc.iac.es}
\and
Gran Telescopio Canarias (GRANTECAN), 38205, San Crist{\'o}bal de La Laguna, Tenerife, Spain
\and
Centre for Astrophysics and Cosmology, Science Institute, University of Iceland, Dunhagi 5, 107, Reykjav\'ik, Iceland
\and
Cosmic Dawn Center (DAWN)
\and
Niels Bohr Institute, University of Copenhagen, Lyngbyvej 2, 2100, Copenhagen \O, Denmark
\and
European Southern Observatory, Alonso de C\'ordova 3107, Vitacura, Casilla 19001, Santiago, Chile
\and
DARK, Niels Bohr Institute, University of Copenhagen, Lyngbyvej 2, 2100, Copenhagen \O, Denmark
\and
Institut d'Astrophysique de Paris, CNRS-UPMC, UMR7095, 98bis Boulevard Arago, 75014, Paris, France
\and
European Southern Observatory, Karl-Schwarzschildstra\ss e 2, 85748, Garching, Germany
}
\authorrunning{Geier et al.}

\date{Received 2019; accepted, 2019}

\abstract{
Damped Lyman-$\alpha$ Absorbers (DLAs) as a class of QSO absorption-line systems are
currently our most important source of detailed information on the cosmic chemical
evolution of galaxies. However, the degree to which this information is biased
by dust remains to be understood. One strategy is to specifically search for QSOs
reddened by metal-rich and dusty foreground absorbers. In this {\it Letter}, we
present the discovery of a $z=2.60$ QSO strongly reddened by dust in an
intervening extremely-strong DLA at $z=2.226$. This QSO was identified through a novel
selection combining the astrometric measurements from ESA's {\it Gaia} satellite with
extent optical and near/mid-infrared photometry. We infer a total neutral
atomic-hydrogen column density of $\log N$(\ion{H}{i}) = $21.95 \pm 0.15$ and a
lower limit on the gas-phase metallicity of [Zn/H] $> -0.96$. This DLA is
also remarkable in exhibiting shielded neutral gas witnessed in \ion{C}{i} and
tentative detections of CO molecular bands. The Spectral Energy Distribution (SED)
of the QSO is well-accounted for by a normal QSO-SED reddened by dust from a DLA
with a 10\%-of-Solar metallicity, dust extinction of $A_V = 0.82 \pm 0.02$\,mag, and
LMC-like extinction curve including the characteristic 2175\,\AA~extinction
feature. Such QSO absorption-line systems have shown to be very rare in previous
surveys, which have mostly revealed sight-lines with low extinction. The present
case therefore suggests that previous samples have under-represented the fraction of
dusty absorbers. Building a complete sample of such systems is needed to assess the
significance of this effect.
}

\maketitle

\section{Introduction}
\label{sec:introduction}

Damped Lyman-$\alpha$ Absorbers (DLAs) have been used to study the cosmic chemical
evolution of galaxies for almost 30 years \citep[e.g.,][]{Pettini1990, Decia2018}.
Huge progress has been made and large samples of well-studied systems are now available
\citep[e.g.,][and references therein]{Berg2016}. From the beginning of this
line of research, it was clear that despite the enormous advantage of the
method in obtaining detailed information on the properties of otherwise
nearly invisible galaxies, the presence of dust will bias the cosmic chemical evolution history
derived this way. This is because DLAs are identified in QSO spectra and the presence
of dust inevitably affects the likelihood of identifying QSOs in most types of QSO selection \citep{FP1989,Pei1991,Pei1999}.

The study of red and/or dust-obscured QSOs also has a long and intricate history \citep[e.g.,][]{Ostriker1984,Webster1995,Benn1998,Warren2000,Glikman06,Glikman2013,Fynbo2013,Banerji2015,Hamann2017,Glikman2018}.
Despite some disagreement among studies about the exact fraction of QSOs missed due
to reddening, it is clear that a substantial number of QSOs are missed when selecting them
only by their optical colours.

A range of studies has attempted to gauge the importance of dust bias in DLA samples.
Direct measurements of the effect of dust reddening have typically found very small amounts of
excess reddening in samples of QSOs with intervening DLAs \citep[][and references 
therein]{Murphy2016}. These studies however used to analyse optically-selected QSO samples.
Radio-selected samples (which are free from dust bias as long as the optical follow-up is 100\% complete) have been built \citep{Ellison2001,Ellison2005,Jorgenson2006}, but these are small and there is room for substantial amounts of missing metals
due to dust-bias \citep{Jorgenson2006, Pontzen2009}.

Our group has specifically aimed at finding examples of QSOs reddened by intervening
dust-rich DLAs using new QSO-selection techniques \citep{Fynbo2013,Krogager2015,Krogager2016b}.
We have found several such systems, which shows that dust bias exists and that our methods succeeds in locating examples of DLAs that have dropped out of optically-selected QSO samples \citep{Krogager2016a,Fynbo2017,Heintz2018a}.
Our ambition is to build complete samples of quasars and, within such samples, identify examples of very metal-rich sight-lines, which are interesting for deep spectroscopic follow-up studies. This is challenging because of the overlap with cool stars for point sources which are red in optical bands. This issue can be alleviated by constructing a complete and (in terms of colour) unbiased sample based on astrometric data, specifically from the 2nd {\it Gaia} data-release \citep[{\it Gaia}-DR2;][]{Gaia2018}, removing sources with significant proper motions, which are not extragalactic \citep{Heintz2015,Heintz2018b}.


\begin{figure}
        \centering 
        \epsfig{file=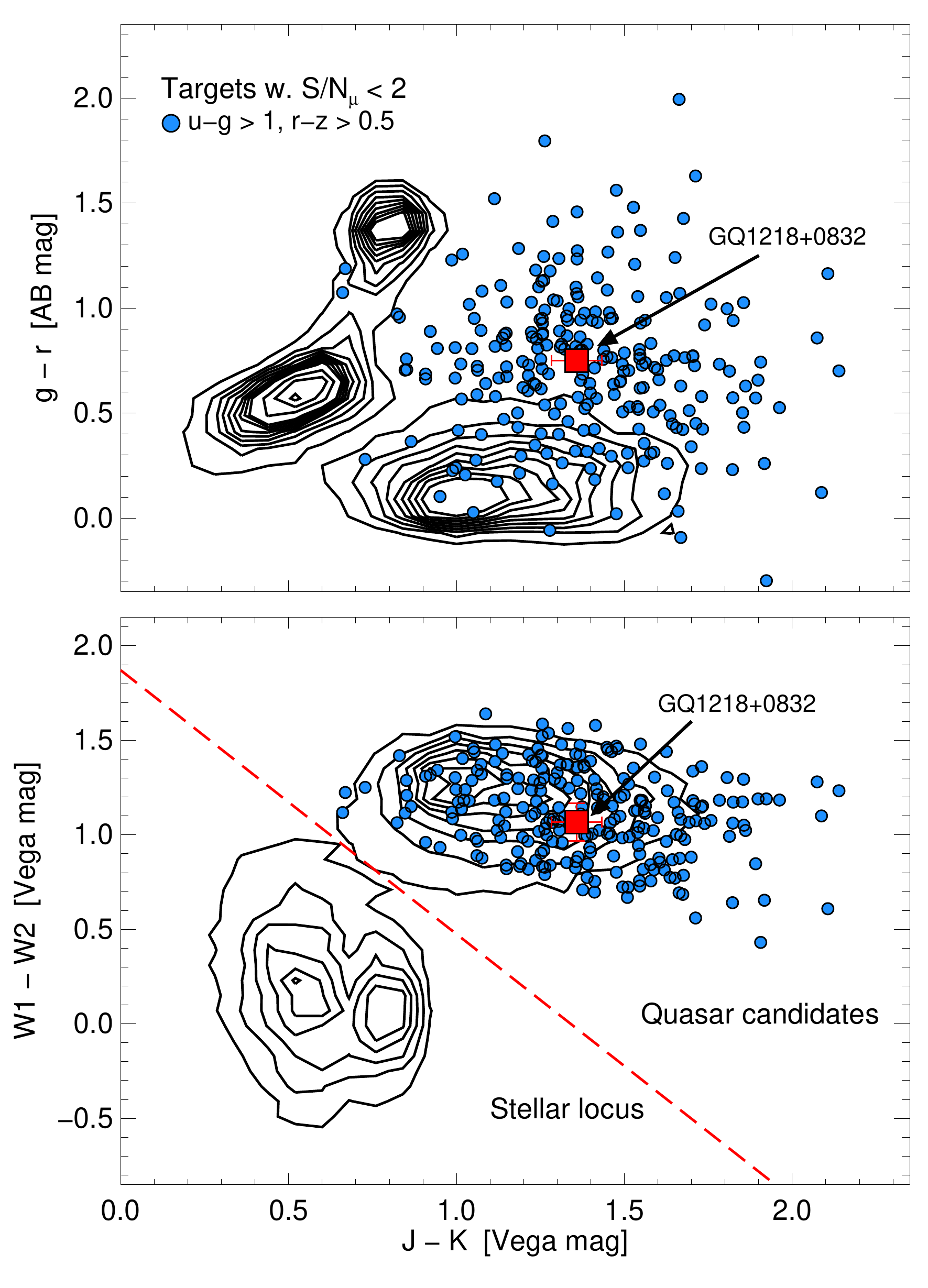,width=9.0cm}
    \caption{Contour plots showing the optical and near-infrared point sources at $b>60^\mathrm{o}$ after the imposed $2\sigma$ zero proper-motion cut. In the upper panel, we display the SDSS $g-r$ vs. UKIDSS $J-K$ colours and, in the bottom panel, the WISE $W1-W2$ vs. UKIDSS $J-K$ colours of the sources in the parent sample. Our near/mid-infrared colour criterion is represented by the red dashed line. The optically-red targets with $u-g > 1$ and $r-z > 0.5$ colours are displayed as blue filled circles and GQ\,1218+0832 is marked with a red square.}
\label{fig:sel}
\end{figure}

This {\it Letter} is organized as follows. In
Sect.~\ref{sec:data}, we present the spectroscopic observations
of a newly-discovered $z=2.60$ QSO with celestial coordinates:
RA $=$ 12:18:30.10, Dec $=$ 08:32:15.5 (J2000.0), called
GQ\,1218$+$0832, reddened by dust in a foreground DLA. In
Sect.~\ref{sec:results}, we present our results on the
characteristics of this DLA, constraining its metallicity and
induced QSO extinction. We draw our conclusions in Sect.~\ref{sec:conc}.

\section{Target selection and observations}    \label{sec:data}

We have built our parent QSO catalog using a novel combination \citep{Heintz2018b} of astrometry from {\it Gaia}-DR2 and photometry
from the optical Sloan Digital Sky Survey data release 12 \citep[SDSS-DR12,][]{Eisenstein2011}, the near-infrared UKIDSS \citep{Warren2007}, and the mid-infrared all-sky WISE mission \citep{Cutri13}.
We confined the sample to a region of the sky of Galactic latitude $b > 60$\,deg, to limit the contamination by Galactic sources. We only considered targets
listed as point sources in both SDSS and UKIDSS and then only included sources with total proper motions $\mu = \sqrt{\mu_{\rm \alpha*}^2 + \mu_{\rm \delta}^2}$ consistent with zero within 2$\sigma$, i.e.,
S/N$_{\mu} = \mu/\mu_{\rm err} < 2$ \citep[following][see Fig.~\ref{fig:sel}]{Heintz2018b}. 
To be as complete as possible, we initially only imposed the zero-proper motion criterion. However, as a significant
number of stars were still present, we imposed the additional near/mid-infrared criteria shown in the bottom panel of Fig.~\ref{fig:sel} to exclude the large majority of star contaminants. Following \citet{Heintz2018a}, we also imposed the optical colour criteria $u-g > 1$ and $r-z > 0.5$ to specifically target dust-reddened QSOs at $z > 2$. 

The object discussed in this {\it Letter}, designated GQ\,1218+0832, was observed as part of a larger spectroscopic campaign. It has $r = 20.35$, $g-r = 1.07$, and
$J-K = 0.42$ (AB mag) and is marked with a red square in Fig.~\ref{fig:sel}. 



Spectroscopic observations were obtained with the OSIRIS instrument at the
Gran Telescopio Canarias (GTC). We gathered spectra using three different grisms in order to better constrain the QSO spectral energy distribution, and intervening metal and \ion{H}{i} Ly$\alpha$ lines. The listed spectral resolution values are based on the measured widths of sky-emission lines. The observing log is given in Table~\ref{tab:log}.

\begin{table}[!t]
\centering
\begin{minipage}{0.5\textwidth}
\centering
\caption{Log of OSIRIS observations.}
\begin{tabular}{lrrcc}
\noalign{\smallskip} \hline \hline \noalign{\smallskip}
Date (UT) & Grism & Resolving & Exp. time & Airmass \\
     &       & Power & [sec]     &         \\
\hline
09/12/2018 & R1000B & 500  & 2$\times$1000 & 1.32--1.22 \\
10/12/2018 & R2500V & 1680 & 1$\times$1500 & 1.92--1.68       \\
11/12/2018 & R2500V & 1680 & 3$\times$1500 & 1.36--1.12 \\
17/12/2018 & R2500R & 1510 & 2$\times$1350 & 1.20--1.12 \\
\noalign{\smallskip} \hline \noalign{\smallskip}
\end{tabular}
\centering
\label{tab:log}
\end{minipage}
\end{table}

\begin{figure*}
\centering
\epsfig{file=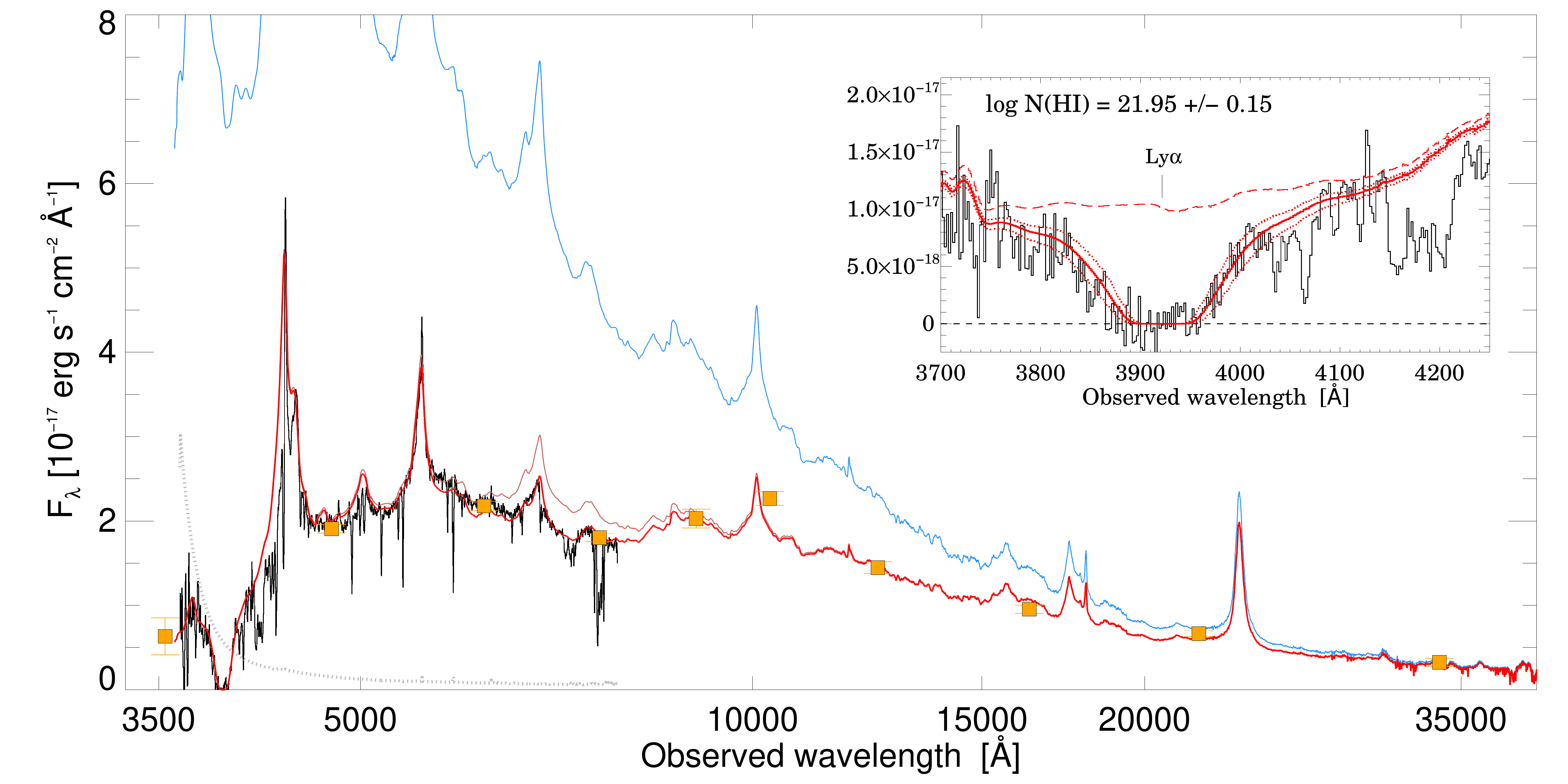,width=17cm}
\caption{The GTC R1000B spectrum of the $z=2.60$ QSO GQ~1218$+$0832 is displayed together with the photometry from SDSS,
UKIDSS, and WISE, in the $u, g, r, i, z, Y, J, H, K_s$, and $W1$ bands \citep{Warren2007, Wright2010,Eisenstein2011}.
Overplotted is the composite QSO spectrum from \citet{Selsing2016} (in blue) and the same composite reddened by $A_V = 0.82\pm 0.02$\,mag assuming LMC-type dust at the redshift of the DLA (in red). Both a model with and without the 2175\,\AA~dust extinction feature are shown. The inset shows a zoom in on the blue part of the spectrum around the \ion{H}{i} Ly$\alpha$ line from the $z=2.2261$ absorber. The red dashed line is the modeled QSO continuum. Also overplotted is the \ion{H}{i} DLA Voigt-profile fit, which is included in the dust-reddened QSO composite spectrum, with $N$(H\,{\sc i}) = $10^{21.95\pm0.15}$ cm$^{-2}$.}
\label{fig:spectrum}
\end{figure*}

The spectroscopic data were reduced using standard IRAF\footnote{IRAF is distributed by the National Optical Astronomy Observatory, which is operated by the Association of Universities for
Research in Astronomy (AURA) under a cooperative agreement with the
National Science Foundation.} procedures. 
The spectra were flux-calibrated using observations of spectro-photometric standard stars observed on each night. Both the spectra and the 
photometric data points were corrected for Galactic extinction using the dust maps from 
\citet{SchlaflyFinkbeiner2011}. All reported wavelengths are in vacuum and 
corrected to the heliocentric rest-frame. To improve the absolute flux 
calibration and correct for slit-loss, we scaled the observed 
spectra to the $r$-band photometry from SDSS.

\section{Results} \label{sec:results}

We display the GTC R1000B spectrum in Fig.~\ref{fig:spectrum}, together with the optical and near/mid-infrared
photometry from SDSS, UKIDSS, and WISE. We identify an extremely strong DLA \citep[hereafter ES-DLA;][]{Noterdaeme14}
at a redshift of $z = 2.2261$ based on a range of associated metal lines (see Table~\ref{tab:EW}).

\subsection{Extinction}

The spectrum of the background QSO is inconspicuous except for its reddening. The fact that the QSO looks "normal" makes it well-suited for the determination of the extinction curve of dust in the intervening ES-DLA.

To derive the amount of extinction, we 
use the composite QSO template from \cite{Selsing2016} including a range of different 
extinction models. We follow the parametrization of \citet{Gordon2003} and fit the extinguished 
spectrum assuming a dust composition similar to the Small and Large Magellanic Clouds (SMC and 
LMC). We find the best-fit to be an LMC-type extinction curve with a visual extinction of 
$A_V = 0.82\pm 0.02$\,mag, including the 2175\,\AA~dust extinction feature. We measure a bump strength of $A_{\rm bump} = 0.38$\,mag \citep[following the definition in][]{Gordon2003}.

\subsection{Absorption lines}

In the inset of Fig.~\ref{fig:spectrum}, we show a zoom-in on the damped Lyman-$\alpha$ absorption line.
To derive the H\,{\sc i} column density of the absorber, we included a DLA in the dust-reddened QSO model
\citep[using the approximation by][]{TepperGarcia2006}.
From the best-fit model, we find $\log N$(H\,{\sc i}) = $21.95\pm 0.15$.

We also detect a range of metal lines associated with the DLA, as
well as another intervening system at $z=2.247$ seen in
\ion{C}{iv} and \ion{Si}{iv}. In Table~\ref{tab:EW}, we list
the equivalent width (EW) and redshift measurements drawn from the
R2500V and R2500R spectra. The average redshift of the system inferred from
the low-ionization lines listed in the table
is $z=2.2261\pm 0.0004$.

We have fitted Voigt profiles to the low-ionization metal lines
and derived the following lower limits on the abundances: [Zn/H] > $-0.96$, [Si/H] > $-1.23$, [Fe/H] > $-1.89$,
and [Cr/H] > $-1.64$, using the Solar photospheric abundances from \citet{Asplund2009}.
Because of the limited resolution of the spectra, these abundances should
be considered cautiously and only treated as lower limits due to
saturation.
Taken at face value, our measurements
indicate an overall metallicity of at least 10\% of Solar
and significant depletion of refractory
elements, e.g., [Fe/Zn$]\sim -1$.


\begin{figure}
\centering
\epsfig{file=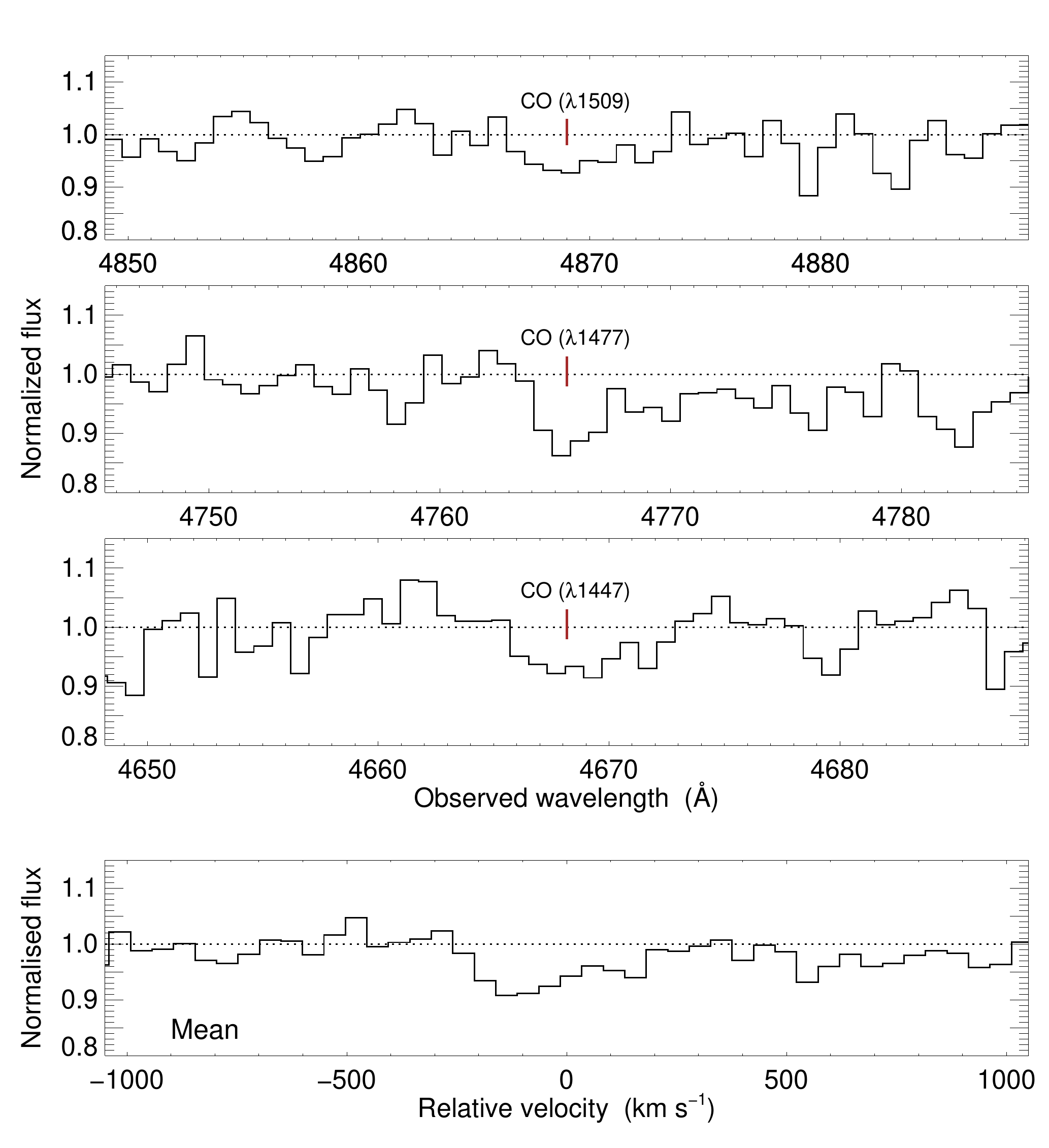,width=8.0cm}
\caption{GTC R2500V spectrum showing tentative detection
of CO A-X bands. In the bottom panel, the geometric mean of the profiles is shown, revealing the line-profile asymmetry.}
\label{fig:co}
\end{figure}

We detect clear absorption signatures of \ion{C}{i} but, due to the limited spectral resolution, the
transitions from the ground state and fine-structure energy levels are blended.
Remarkably, there are consistent features also at the expected locations of CO molecular bands (see Fig.~\ref{fig:co}). A deeper, higher resolution spectrum is needed
to confirm the presence of these lines.

\begin{table}[!htbp]
\centering
\caption{Absorption lines from the $z=2.2261\pm0.0004$ DLA
detected in the GTC R2500V and R2500R spectra.}
\begin{minipage}{0.50\textwidth}
\centering
\begin{tabular}{lcc}
\noalign{\smallskip} \hline \hline \noalign{\smallskip}
Transition & \emph{EW$_\mathrm{r}$} & Redshift \\
           & [\AA]                     &          \\
\hline
\ion{Si}{iv}\,$\lambda$\,1393      & $0.41\pm0.11$ & 2.2262 \\
\ion{Si}{iv}\,$\lambda$\,1402$^a$  & $0.53\pm0.09$ & 2.2262 \\
CO\,$\lambda$\,1447  & $0.20\pm0.11$ & 2.2265 \\
CO\,$\lambda$\,1477  & $0.09\pm0.06$ & 2.2265 \\
CO\,$\lambda$\,1509  & $0.09\pm0.07$ & 2.2265 \\
\ion{Si}{ii}\,$\lambda$\,1526  & $1.21\pm0.07$ & 2.2257 \\
\ion{C}{iv}\,$\lambda$\,1548  & $0.48\pm0.05$ & 2.2266 \\
\ion{C}{iv}\,$\lambda$\,1550  & $0.33\pm0.05$ & 2.2255 \\
\ion{C}{iv}\,$\lambda$\,1548  & $0.46\pm0.04$ & 2.2474 \\
\ion{C}{i}\,$\lambda$\,1560$^a$  & $0.54\pm0.04$ & 2.2265 \\
\ion{Fe}{ii}\,$\lambda$\,1608  & $1.10\pm0.06$ & 2.2257 \\
\ion{Fe}{ii}\,$\lambda$\,1611  & $0.11\pm0.04$ & 2.2256 \\
\ion{C}{i}\,$\lambda$\,1656  & $0.58\pm0.06$ & 2.2265 \\
\ion{Al}{ii}\,$\lambda$\,1670  & $1.43\pm0.05$ & 2.2260 \\
\ion{Ni}{ii}\,$\lambda$\,1709  & $0.15\pm0.03$ & 2.2259 \\
\ion{Ni}{ii}\,$\lambda$\,1741  & $0.22\pm0.03$ & 2.2260 \\
\ion{Ni}{ii}\,$\lambda$\,1751  & $0.19\pm0.04$ & 2.2263 \\
\ion{Si}{ii}\,$\lambda$\,1808  & $0.67\pm0.04$ & 2.2260 \\
\ion{Al}{iii}\,$\lambda$\,1854  & $0.35\pm0.05$ & 2.2262 \\
\ion{Al}{iii}\,$\lambda$\,1862  & $0.20\pm0.05$ & 2.2266 \\
\ion{Ti}{ii}\,$\lambda$\,1910  & $0.13\pm0.04$ & 2.2261 \\
\ion{Zn/Cr}{ii}\,$\lambda$\,2026  & $0.51\pm0.04$ & 2.2268 \\
\ion{Cr}{ii}\,$\lambda$\,2056  & $0.23\pm0.04$ & 2.2259 \\
\ion{Zn/Cr}{ii}\,$\lambda$\,2062  & $0.45\pm0.04$ & 2.2257 \\
\ion{Cr}{ii}\,$\lambda$\,2066  & $0.14\pm0.04$ & 2.2265 \\
\ion{Fe}{ii}\,$\lambda$\,2249  & $0.17\pm0.04$ & 2.2261 \\
\ion{Fe}{ii}\,$\lambda$\,2260  & $0.36\pm0.04$ & 2.2265 \\
\ion{Fe}{ii}\,$\lambda$\,2344  & $1.93\pm0.05$ & 2.2261 \\
\noalign{\smallskip} \hline \noalign{\smallskip}
\end{tabular}
\end{minipage}
\flushleft{$^a$ Blended with \ion{C}{iv}\,$\lambda$\,1550 at $z_{\rm abs}=2.2474$.}
\label{tab:EW}
\end{table}


\section{Discussion and conclusions}
\label{sec:conc}

\begin{figure}
\centering
\epsfig{file=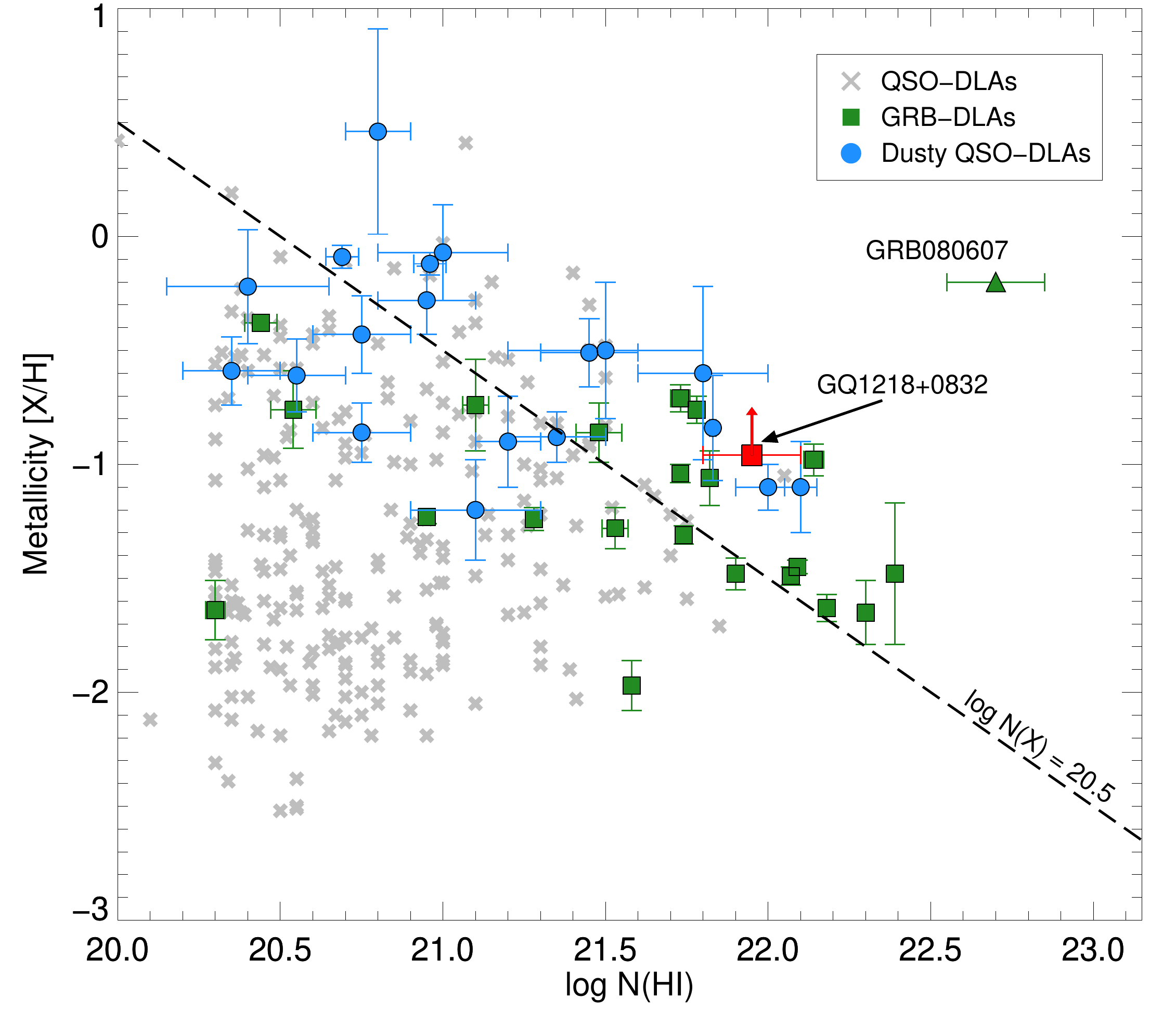,width=9.0cm}
\caption{Metallicity versus \ion{H}{i} column density. Typical QSO-DLAs with metallicities from the literature \citep{Decia2018} are displayed as grey crosses. The sample of GRB-DLAs from \citet{Bolmer2018} is shown with green squares, and the remarkable GRB\,080607 presented in \citet{Prochaska2009} is marked with a green upward-facing triangle. A sample of dusty \citep[$A_V > 0.1$\,mag, compiled by][]{Heintz2018a} QSO-DLAs is shown with blue circles.  The dashed line corresponds to a fixed metal column density, i.e. $\log N(X) = \log N(\ion{H}{i}) + [X/H] = 20.5$ \citep[see also][]{Boisse1998}. The presently-studied ES-DLA towards GQ\,1218+0832 is marked with a red square, indicating the lower limit on the metallicity. It is among the most metal-rich ES-DLAs discovered to date.}.
\label{fig:himet}
\end{figure}

In Fig.~\ref{fig:himet}, we compare the metallicity and \ion{H}{i} column density of the
presently-studied system to the general population of
QSO-DLAs from \citet{Decia2018}, GRB-DLAs from \citet{Bolmer2018}, the extreme case of GRB~080607 from \citealt{Prochaska2009}, and
the compiled sample of dusty ($A_V > 0.1$\,mag) DLAs from \citet{Heintz2018a}, including the recent discovery by \citet{Ranjan2018}.
Compared to other QSO-DLAs, the absorption system towards GQ\,1218+0832 is remarkable in several ways. Firstly, such a high \ion{H}{i} column density is very rare among the general population of QSO absorbers \citep{Bird2017}. Even within the population of extremely strong QSO-DLAs, this system has higher-than-average metal-line strengths, depletion factors, and H\,{\sc i} column density \citep{Noterdaeme14}.
Its \ion{H}{i} column density is more similar to the column densities found in GRB-DLAs \citep[see also][]{Tanvir2019}. This DLA is also unusual given the presence of \ion{C}{i} and possibly CO absorption. The incidence of \ion{C}{i} absorbers with $W_r({\lambda 1560})>0.4~{\AA}$ is about one-hundred times less than that of QSO-DLAs \citep{Ledoux2015}. The presence of strong \ion{C}{i} lines is an indicator of high metallicity, diffuse molecular gas, and \citet{Noterdaeme2018} found that all such systems present H$_2$ and a significant fraction of them exhibit CO at a detectable level. This suggests that CO lines are likely to be detected in this system as well, in agreement with the presence of consistent dips in
the spectrum. 
This system provides further evidence that ES-DLAs (with $N$(\ion{H}{i}) > $10^{21.7}$ cm$^{-2}$) lie in the transition regime where \ion{H}{i} is being converted into H$_2$ \citep{Schaye2001,Noterdaeme2015,Balashev2018}.


This case, in terms of extinction and molecular content, is the most
remarkable example so far of a QSO missed by traditional optical-colour
selections due to dust in an intervening galaxy as predicted by \citet{FP1989}.

How many quasars would be missed if they had a DLA in their line-of-sight with
the same level of extinction as in the present case? The answer to this question is
complex, as the effect of dust depends both on the redshift of the quasar,
the redshift of the DLA, and the colour distribution of stellar populations. In this survey, we find that the fraction of candidate quasars with the same red optical colours as the system presented here is $\sim 1\%$, compared to the full set of quasars identified based on the astrometric and near/mid-infrared selection criteria detailed in Sect.~\ref{sec:data}.
\citet{Krogager2019} have carried out a comprehensive analysis of the effect of
dust in DLAs and found that whereas the fraction of quasars dropping out of selections
is small, the effect on the derived mass density of neutral hydrogen and metals is much larger, up to a factor of 2 and 5 at $z\sim 2.2$, respectively \citep[see also][]{Pontzen2009}.

There is little doubt that more such systems are yet to be found. An
important objective of future work will be to identify them. Another major
endeavour will be to use the information of their incidence rate to build up
a more complete picture of cosmic chemical evolution.

\begin{acknowledgements}
We thank the anonymous referee for a
constructive report that
helped improve the manuscript on important points. This work is based on 
observations carried out on the island of
La Palma with the Gran Telescopio Canarias (GTC), installed at
the Spanish Observatorio del Roque de los Muchachos belonging to
the Instituto de Astrofísica de Canarias. The Cosmic Dawn Center
is funded by the DNRF. KEH and PJ acknowledge support from a
Project Grant (162948--051) of The Icelandic Research Fund.
\end{acknowledgements}

\bibliographystyle{aa}
\bibliography{ref}

\object{GQ\,1218+0832}
\end{document}